\documentclass{article}

\usepackage{amssymb}
\usepackage{amsmath}
\usepackage{float}
\usepackage{cite}
\usepackage{graphics}
\usepackage{epsfig}

\usepackage{graphicx}

\begin{document}


\noindent \noindent \textbf{} \textbf{\centerline{\LARGE A Higher Dimensional Cosmological Model }}\\

\noindent \noindent \textbf{} \textbf{\centerline{\LARGE for the Search of Dark Energy Source}}\\

\noindent \textbf{}

\noindent \textbf{}

\noindent \noindent \textbf{} \textbf{\centerline{$^1$Pheiroijam Suranjoy Singh, $^2$Kangujam Priyokumar Singh }}\\

\noindent \noindent {\centerline{  $^1$$^,$$^2$Department of Mathematical Sciences,}}

\noindent \noindent {\centerline{  Bodoland University, Kokrajhar, Assam-783370, India}}

\noindent \noindent {\centerline{  $^2$Department of Mathematics,}}

\noindent \noindent {\centerline{  Manipur University, Imphal, Manipur-795003, India}}\\

\noindent \noindent {\centerline{ $^1$E-mail : surphei@yahoo.com}}

\noindent \noindent {\centerline{ $^2$E-mail : pk\textunderscore{mathematics@yahoo.co.in}}}\\

\noindent \textbf{}

\noindent

\noindent

\begin{abstract}
With due consideration of reasonable cosmological assumptions within the limit of the present cosmological scenario, we have analysed a spherically symmetric metric in 5D setting within the framework of Lyra manifold. The model universe is predicted to be a DE model, dominated by vacuum energy. The model  represents an oscillating model, each cycle evolving with a big bang and ending at a big crunch, undergoing a series of bounces. The universe is isotropic and undergoes super-exponential expansion. The value of Hubble's parameter is measured to be $H=67.0691$ which is very close to $H_0=67.36\pm0.54kms^{-1}Mpc^{-1}$, the value estimated by the latest Planck 2018 result. A detailed discussion on the cosmological parameters obtained is also presented with graphs. 
\end{abstract}

\section{Introduction}

The enigmatic dark energy (DE) has gained the reputation of being one of the most discussed topics of paramount importance in cosmology since its profound discovery in 1998 \cite{1,2}. Its property with huge negative pressure with repulsive gravitation causing the universe to expand at an expedited rate still remains a mystery. It is believed to uniformly permeate throughout the space and vary slowly or almost consistent with time \cite{3,4,5,6}. Cosmologists around the world have invested tremendous scientific efforts with a strong focus to hunt its origin and are still scrabbling for a perfect answer. Different authors have put forward their own versions of the answer with convincing evidences in support. Some worth mentioning studies on the search of the root of DE which have not escaped our attention in the past few years are briefly discussed below.

In \cite{7}, the authors assert that emergent D-instanton might lead us to the root of DE. The analysis of the twenty years old history of DE and the current status can be seen in \cite{8}. In \cite{9}, the authors investigate the evolution of the DE using a non-parametric Bayesian approach in the light of the latest observation. In \cite{10}, the author claims that vacuum condensate can indicate us the origin of DE. A study on neutrino mixing as the origin of DE is presented in \cite{11}. In \cite{12}, the explanation for DE with pure quantum mechanical method is presented. In \cite{13}, it is mentioned that DE emerges due to condensation of fermions formed during the early evolution. The explanation of a physical mechanism as a source of DE can be seen in \cite{14}. A cosmological model involving an antineutrino star is proposed in \cite{15} in an attempt to find the origin of DE. In \cite{16}, DE  is obtained from the violation of energy conservation. A unified dark fluid is obtained as a source of DE in \cite{17}. Lastly, in \cite{18}, the author claims that the presence of particle with imaginary energy density can lead us to the source of DE. 

From literatures and observations \cite{4,19,20,21,22,23,24,25}, it is obvious that the massive universe is dominated by the mystic DE with negative pressure and positive energy density. This qualifies DE a completely irony of nature as the dominating component is also the least explored. So, there a lot more hidden physics behind this dark entity yet to be discovered. Contradicting to the condition of positive energy density, it is surprising that many authors has come up with the notion of negative energy density (NED) with convincig and fascinating arguments in support.  In \cite{4}, the authors predicted a condition in which NED is possible only if the DE is in the form of vacuum energy. Besides defying the energy conditions of GR, NED also disobeys the second law of thermodynamics \cite{26}. However, the condition should be solely obeyed on a large scale or on a mean calculation, thereby neglecting the probable violation on a small scale or for a short duration, in relativity \cite{27,28,29,30,31,32,33,34,35}. Hence, in the initial epoch, if there were circumstance of defiance for a short duration measured against the present age which is estimated to be $13.825\pm0.037$ Gyr by the latest Planck 2018 result \cite{36}, it will remain as an important part in the course of evolution. In \cite{37}, under certain conditions, a repelling negative gravitational pressure can be seen with NED. It further mention of a repelling negative phantom energy with NED. Energy density assuming negative value with equation of state parameter (EoS) $\omega<-1$ can be found in \cite{38}. In \cite{39}, the author asserts that the universe evolves by inflation when the coupled fluid has NED in the initial epoch. The authors in \cite{40} discuss negative vacuum energy density in Rainbow Gravity. According to \cite{41}, the introduction of quantized matter field with NED to energy momentum tensor might by pass cosmological singularity. In \cite{42}, the author investigate models which evolved with NED in the infinite past. In \cite{43}, NED is discussed where the authors assert that their models evolve with a bounce. The authors continued that there might be bounces in the future too. Lastly, an accelerating universe with NED is studied in \cite{44}.

The sessions of the Prussian Academy of Science during the four Thursdays of the month of November 1915 can be marked as the most memorable moments in the life of the great Einstein. On 4\textsuperscript{th}, 11\textsuperscript{th}, 18\textsuperscript{th} and 25\textsuperscript{th} of the month, he presented four of his notable communications \cite{45,46,47,48} at the sessions, which led to the foundation of GR. Since then, a number of authors have been exploring gravitation in different settings. In Weyl's work of 1918 \cite{49}, we can witness the first trial to extend GR with the aim of bringing together gravitation and electromagnetism geometrically. Similar to that of Weyl, Lyra's modification \cite{50} by proposing a gauge function into the structureless manifold provides one of the well appreciated alternate or modified theories of gravitation. Lyra's work are further extended by other well known aothors \cite{51,52,53,54,55}. Cosmologists choose to opt alternate or modified theories of gravitation in order to precisely understand the underlying mechanism of the late time expedited expansion of the universe. Many other authors too have secceded in developing fascinating and worth appreciating modified theories which have served the purpose of explaining the expanding paradigm in a quite convincing way \cite{56,57,58,59,60,61,62,63,64}.

During the past few years, there has been an increasing interest among cosmologists to study the ambiguous DE paired with Lyra Manifold (LM). Recently, in \cite{65}, the authors investigate the existence of Lyra’s cosmology with interaction of normal matter and DE. In \cite{66}, we can witness a DE model in a LM which proves that the expansion paradigm can be illustrated in the absence of a negative pressure energy component. It is further mentioned that DE is naturally of geometrical origin. The investigation of a two component DE model in LM can be seen in \cite{67}. In \cite{68}, we can find a cosmological model of anisotropic DE paired with LM  which is in consonant with the present observation. The study of of a magnetized DE model in Lyra setting can be seen in \cite{69}. A discussion on the effect of DE on model with linear varying deceleration parameter in LM  can be found in \cite{70}. In \cite{71}, the authors present the isotropization of DE distribution in LM . In \cite{72}, Kaluza-Klein DE model is studied in LM thereby obtaining an exponentially expanding universe. In \cite{73}, we can see a DE model in LM where constant deceleration parameter is assumed. Brans–Dicke scalar field as a DE candidate in LM is illustrated in \cite{74}. Lastly, in \cite{75}, a DE model with quadratic equation of state is presented in the framework of LM.

The chance of space-time having more than 4D has captivated many authors. This has ignited a spark of interest among cosmologists and theorological physicists so that, in the past few decades, there has been a trend among authors opting to choose higher dimensional space-time to study cosmology. Higher dimensional model was introduced in \cite{76,77} in an attempt to unify gravity with electromagnetism. Higher dimensional model can be regarded as a tool to illustrate the late time expedited expanding paradigm \cite{78}. Investigation of higher dimensional space-time can be regarded as a task of paramount importance as the universe might have come across a higher dimensional era during the initial epoch \cite{79}. In \cite{80,81}, it is mentioned that extra dimensions generate huge amount of entropy which gives possible solution to flatness and horizon problem. In \cite{82}, it is asserted that the detection of a time varying fundamental constants can possibly show us the proof for extra dimensions. Since we are living in a 4D space-time, the hidden extra dimension in 5D is highly likely to be associated with the invisible DM and DE \cite{83}. Few of the worth mentioning works on higher dimensional space-time during the last few years can be seen in \cite{20,84,85,86,87,88,89,90,91,93}.

Keeping in mind the above notable works by different authors, we have analysed a spherically symmetric metric in 5D setting within the framework of LM with the aim of predicting a possible source of DE. Here, we observe the field equations with due consideration of reasonable cosmological assumptions within the limit of the present cosmological scenario. The paper has been structured into sections. In Sect. 2, in addition to obtaining the solutions of the field equations, the cosmological parameters are also solved. In Sect. 3, the physical and kinematical aspects of our model are discussed with graphs. Considering everything, a closing remark is presented in Sect. 4.

\section{Formulation of problem with solutions}

The five-dimensional spherically symmetric metric is given by

\begin{equation} 
ds^{2} =dt^{2} -e^{\mu } \left(dr^{2} +r^{2} d\theta ^{2} +r^{2} \sin ^{2} \theta d\phi ^{2} \right)-e^{\delta } d\upsilon ^{2}       
\end{equation}

\noindent where $\mu=\mu(t) $ and $\delta=\delta(t) $ are cosmic scale factors.\\

The modified Eintein's field equations in Lyra geometry appear in the form

\begin{equation} 
R_{i\, j} -\frac{1}{2} g_{i\, j} R+\frac{3}{2} \varphi _{i} \varphi _{j} -\frac{3}{4} g_{i\, j} \varphi _{k} \varphi ^{k} =-T_{i\, j}       
\end{equation}

\noindent where $\varphi _{i} $ is the displacement vector and other symbols have their usual meaning as in Riemannian geometry. The displacement vector $\varphi _{i} $ takes the time dependent form 

\begin{equation}
\varphi _{i} =(\beta (t),\, \, 0,\, \, 0,\, \, 0,\, \, 0)           
\end{equation}

If $\varphi _{i} $ is time independent i.e. constant, then it acts as the cosmological constant. However, assuming displacement vector field as a constant is just for convenience sake without any scientific reason \cite{93}.   
 
The energy momentum tensor $T_{i\, j} $, considered as a perfect fluid, in the co-moving coordinates is given by 

\begin{equation} 
T_{i\, j} =\left(\rho +p\right)u_{i} u_{j} -pg_{i\, j}        
\end{equation}

\noindent where $\rho $ and $p$ respectively represent the energy density and isotropic pressure of the matter source. The five velocity vector $u^{i}$ satisfies 
\begin{equation} 
u^{i} u_{i} =1,\, \, u^{i} u_{j} =0           
\end{equation}

Now, the surviving field equations are obtained as follows

\begin{equation} 
\frac{3}{4} \left(\dot{\mu }^{2} +\dot{\mu }\dot{\delta }-\beta ^{2} \right)=\rho         
\end{equation}

\begin{equation} 
\ddot{\mu }+\frac{3}{4} \dot{\mu }^{2} +\frac{\ddot{\delta }}{2} +\frac{\dot{\delta }^{2} }{4} +\frac{\dot{\mu }\dot{\delta }}{2} +\frac{3}{4} \beta ^{2} =-p      
\end{equation}

\begin{equation} 
\frac{3}{2} \left(\ddot{\mu }+\dot{\mu }^{2} \right)+\frac{3}{4} \beta ^{2} =-p         
\end{equation}

\noindent where an overhead dot represents differentiation w.r.t. $t$. \\

From continuity equation, we have

\begin{equation} 
\dot{\rho }+\frac{3}{2} \dot{\beta }\beta +3H\left(\rho +p+\frac{3}{2} \beta ^{2} \right)=0          
\end{equation}

From \cite{94}, assuming that $\beta $ and $\rho $ are independent without any interaction, Eq. (9) can be separately written as 

\begin{equation}
\dot{\rho }+3H\left(\rho +p\right)=0            
\end{equation}

\begin{equation}
\dot{\beta }\beta +3H\beta ^{2} =0           
\end{equation}

\noindent where $H$ is the Hubble's parameter.\\

From Eqs. (7) and (8), the expression for cosmic scale factors are obtained as

\begin{equation}
\mu =l-\log \left(k-t\right)^{\frac{2}{3} }            
\end{equation}

\begin{equation}
\delta =m-\log (k-t)^{\frac{2}{3} }             
\end{equation}

\noindent where $l,\, \, m,\, \, k$ are arbitrary constants.\\

Now, we obtain the expression for the cosmological parameters as follows.\\

Spatial volume:  
\begin{equation}
v=e^{\frac{3\mu+\delta}{2}} =e^{\frac{3l+m}{2} } \left(k-t\right)^{-\frac{4}{3} } 
\end{equation}

Scale factor:  
\begin{equation}
a(t)=v^{\frac{1}{4}}=e^{\frac{3l+m}{8}} \left(k-t\right)^{-\frac{1}{3} } 
\end{equation} 

Scalar expansion: 
\begin{equation}
\Theta =u_{;\, j}^{i}= \frac{3\dot{\mu} }{2}+  \frac{\dot{\delta } }{2}= \frac{4}{3} \left(k-t\right)^{-1} 
\end{equation}

Hubble's parameter: 
\begin{equation}
H=\frac{\Theta}{4}=\frac{1}{3} \left(k-t\right)^{-1} 
\end{equation}

Deceleration parameter: 
\begin{equation}
q=\frac{d}{dt}\left(\frac{1}{H}\right)-1=-4       
\end{equation}

With $\Delta H_i=H_i-H, (i=1, 2, 3, 4)$ representing the directional Hubble's parameters, anisotropic parameter $A_h$ is defined as
\begin{equation}
A_h=\frac{1}{4}\sum_{i=1}^{4} \left(\frac{\Delta H_i}{H}\right)^2 =0
\end{equation}

Shear Scalar: 
\begin{equation}
\sigma ^{2} =\frac{1}{2}\sigma_{i\,j}\sigma ^{i\,j}=\frac{1}{2}\sum_{i=1}^{4} \left( H_i^2-4H\right)=\frac{2}{9} \left(1-\frac{1}{k-t} \right)^{2} 
\end{equation}        

From Eqs. (11) and (17), the expression for displacement vector is obtained as

\begin{equation} 
\beta =d\, \left(t-k\right)           
\end{equation}

\noindent where $d$ is an arbitrary constant.\\

From Eqs. (6), (12), (13) and (21), the expression for energy density is obtained as

\begin{equation}
\rho =\frac{3}{4} \left(\frac{8-\left(\, 3d\, \left(k-t\right)^{2} \, \right)^{2} }{9\left(k-t\right)^{2} } \right)         
\end{equation}

From Eqs. (8), (12) and (13), the expression of pressure is obtained as  

\begin{equation}
p=-\frac{8\left(1+\left(k-t\right)^{2} \right)+\left(3d\left(k-t\right)^{3} \right)^{2} }{12(k-t)^{4} }           
\end{equation}

\section{Discussion}

For convenience sake and to obtain realistic results, specific values of the constants are chosen i.e., $l=m=1,  d=-1, k=3.45497$ and the variations of some of the physical parameters with cosmic time \textit{t} are provided as figures in this section. A scale of 1 Unit = 4 Gyr is taken along the time axis of each graph so that the point $t=3.45$ corresponds to 13.8 Gyr which align with $13.825\pm0.037$ Gyr, the present age of the universe estimated by the latest Planck 2018 result \cite{36}.
\begin{figure}[H]
\centerline{\psfig{file=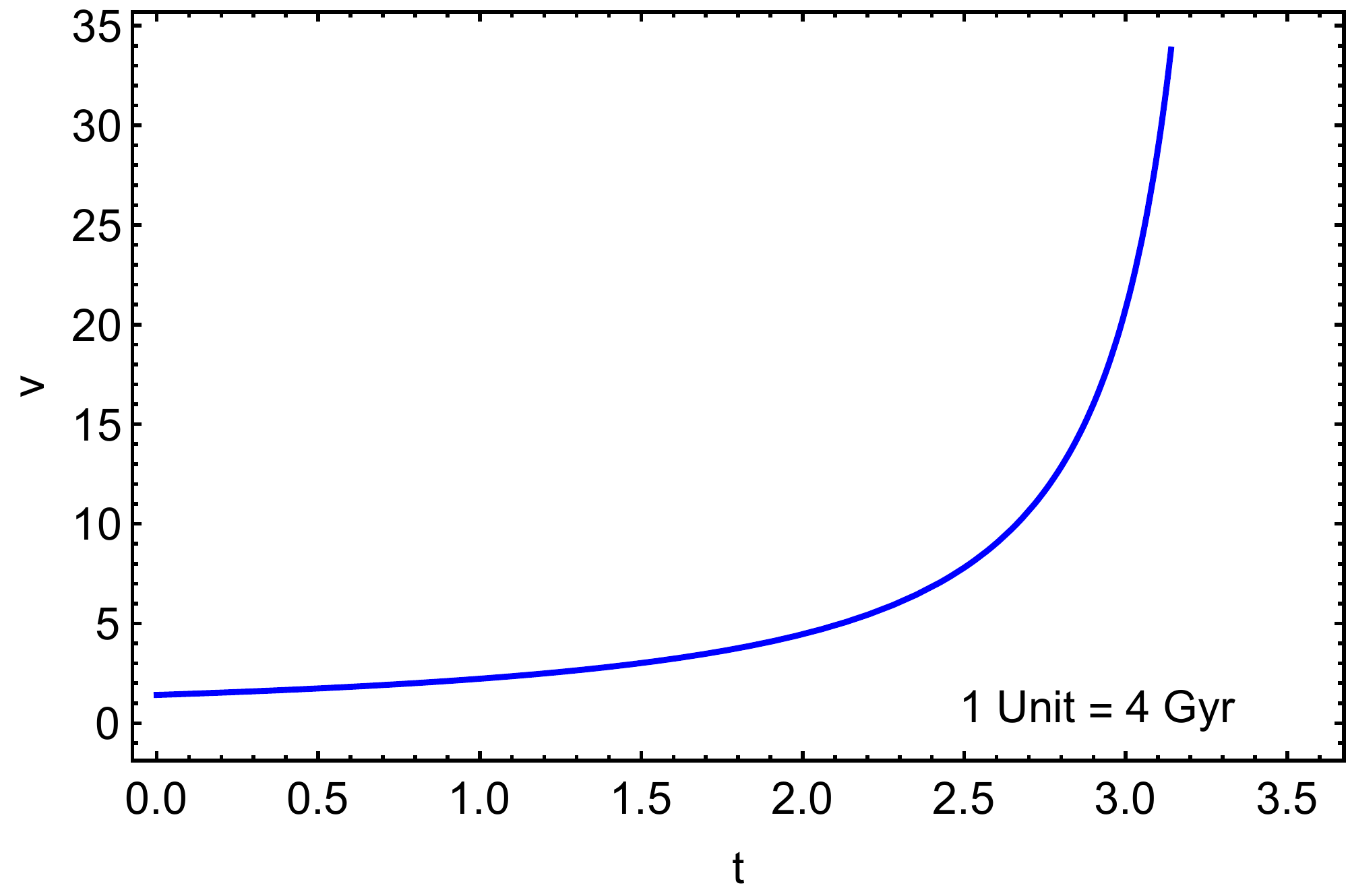,width=2.5in}}
\vspace*{8pt}
\caption{Variation of spatial volume $v$ with time \textit{t} when $l=m=1, k=3.45497$ showing its increasing nature throughout the evolution.\label{fig1}}
\end{figure}

\begin{figure}[H]
\centerline{\psfig{file=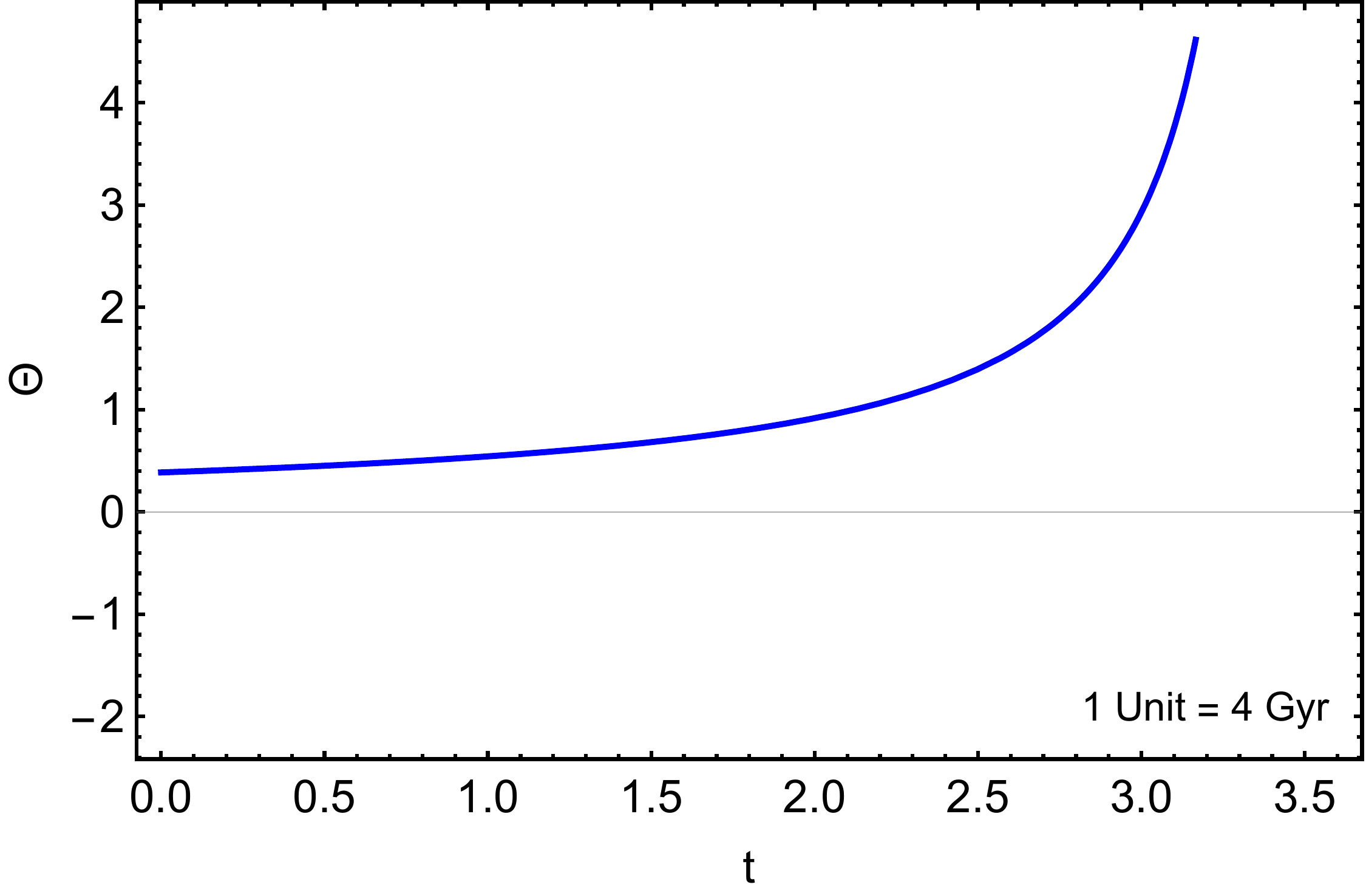,width=2.5in}}
\vspace*{8pt}
\caption{Variation of scalar expansion $\Theta$ with time \textit{t} when $l=m=1, k=3.45497$ showing its increasing nature throughout the evolution.\label{fig1}}
\end{figure}

\begin{figure}[H]
\centerline{\psfig{file=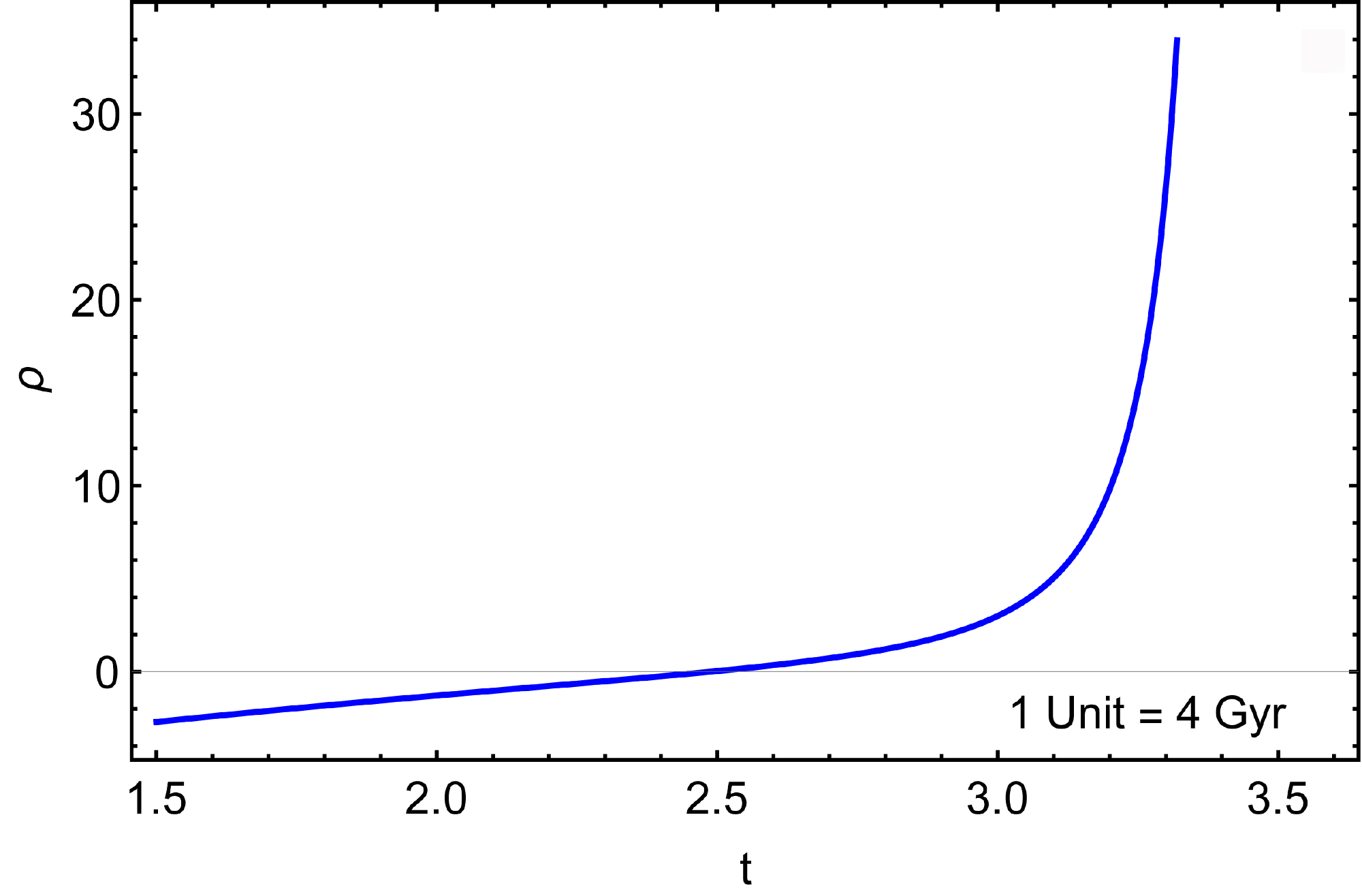,width=2.5in}}
\vspace*{8pt}
\caption{Variation of energy density $\rho$ with time \textit{t} when $l=m=1, k=3.45497$ showing its transition from being negative to positive during evolution.\label{fig1}}
\end{figure}

\begin{figure}[H]
\centerline{\psfig{file=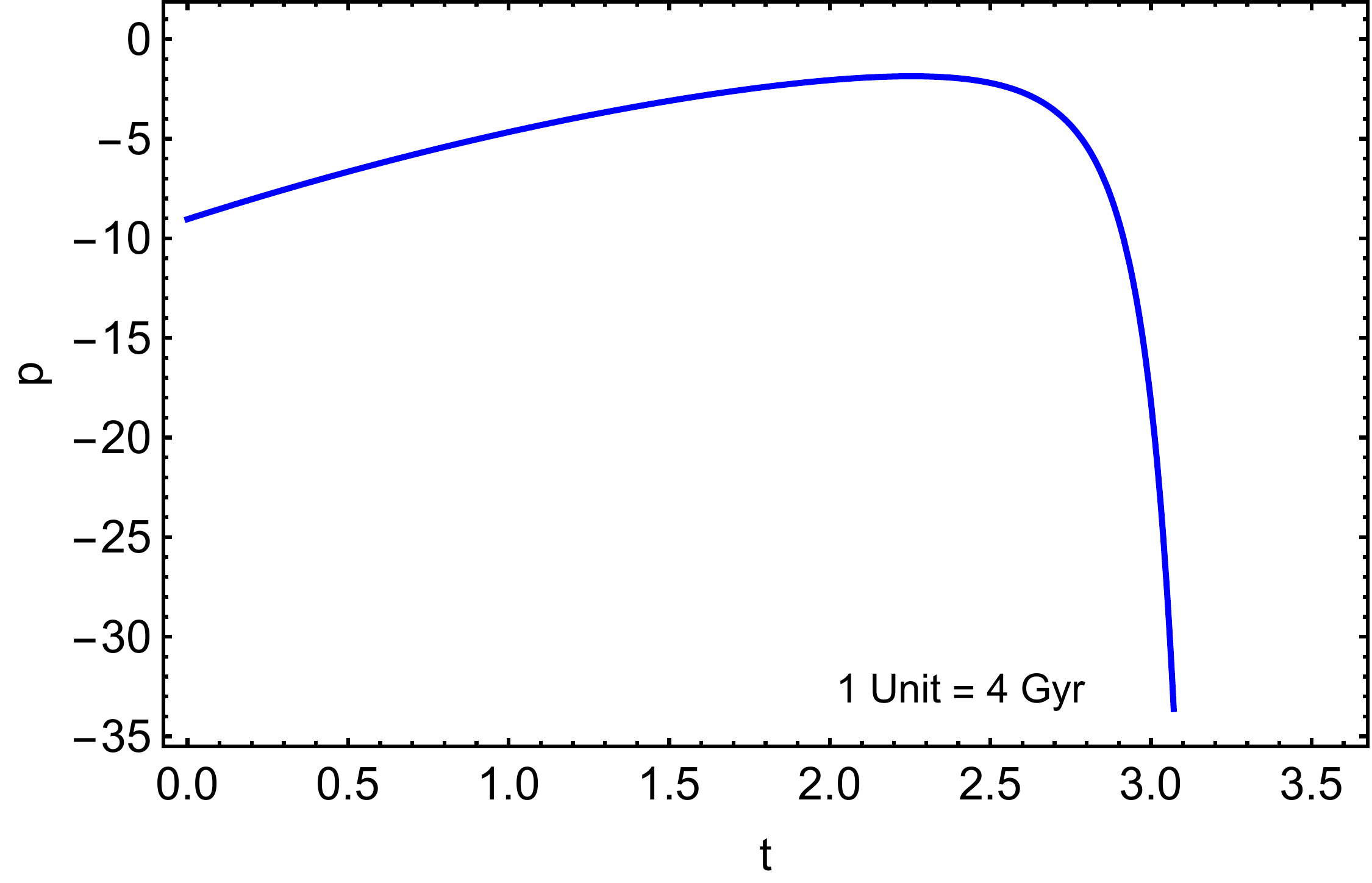,width=2.5in}}
\vspace*{8pt}
\caption{Variation of pressure $p$ with time \textit{t} when $l=m=1, k=3.45497$ showing its negative nature all through.\label{fig1}}
\end{figure}

\begin{figure}[H]
\centerline{\psfig{file=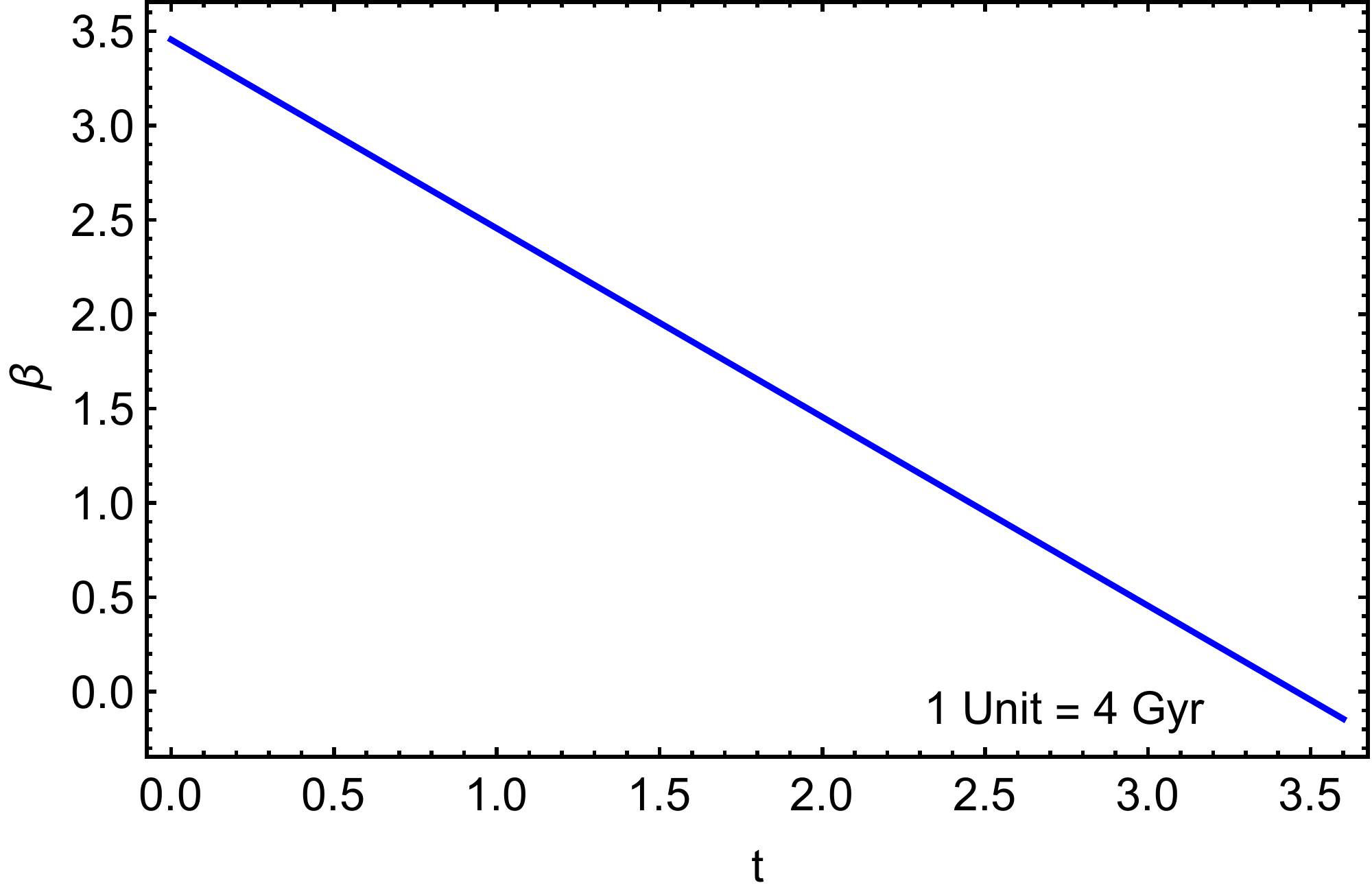,width=2.5in}}
\vspace*{8pt}
\caption{Variation of displacement vector $\beta$ with time \textit{t} when $d=-1, k=3.45497$ showing its decreasing nature.\label{fig1}}
\end{figure}

Figures 1 and 2 can be regarded as the perfect evidences of the present spacial expansion of the universe at an expedited rate. From Figs. 3 and 4, we can witness a transition of the energy density $\rho$ of the model universe from being negative to positive during the course of evolution whereas the pressure $p$ of the model is negative all through. In short, the universe expands at an expedited rate with $\rho$ and $p$ both negative. This negative $p$ can be regarded as the indication of the presence of DE. In this scenario, we can predict that DE in the form of vacuum energy is dominating the model, as mentioned in \cite{4}, NED is possible only if the DE is in the form of vacuum energy. When $t\rightarrow\infty$, both $v$ and $\theta\rightarrow 0$ showing that, in the far future, the expanding phenomenon will cease, the universe will be dominated by gravity, resulting to collapse and ultimately ending at the big crunch singularity. This may be supprted by the fact that DE density may decrease faster than matter leading DE to vanish at $t\rightarrow\infty$ \cite{6}. Additionally, when $t\rightarrow\infty$, $\rho$ again starts to become negative. In this condition, due to the presence of NED, from \cite{43}, we can assume that the model universe represents an oscillating model, each cycle evolving with a big bang and ending at a big crunch, undergoing a series of bounces. Additionally, from Fig. 5, it is clear that the displacement vector $\beta$ is a decreasing function of time. Here, we can assert that $\beta$ acts as the time dependent cosmological constant \cite{1,2,51,95}. Hence, it is fascinating to observe that LM itself can be regarded as a DE model.

\begin{figure}[H]
\centerline{\psfig{file=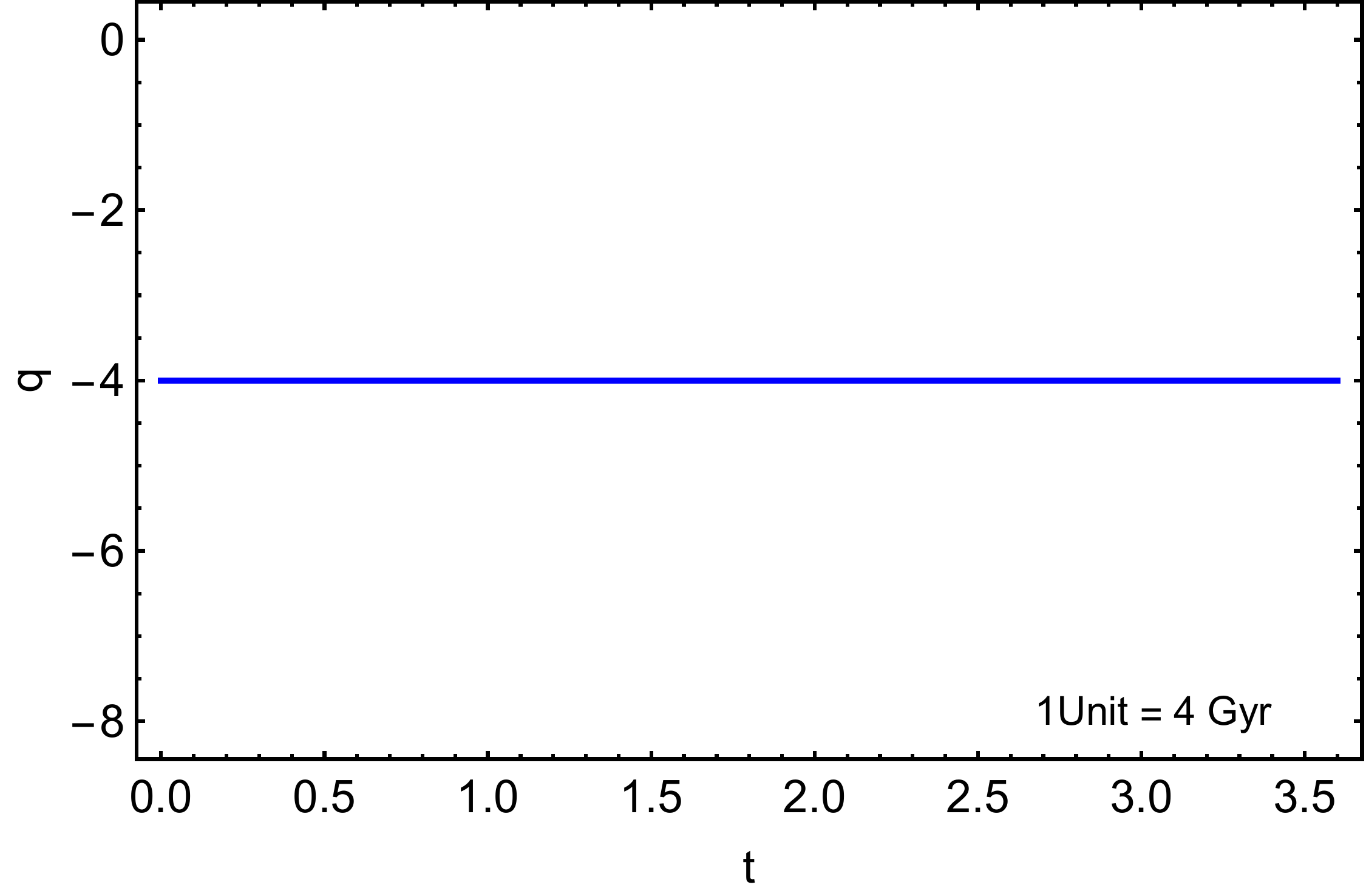,width=2.5in}}
\vspace*{8pt}
\caption{Variation of deceleration parameter $q$ with time \textit{t} when $l=m=1, k=3.45497$ showing that it is a negative constant -4 all through.\label{fig1}}
\end{figure}

\begin{figure}[H]
\centerline{\psfig{file=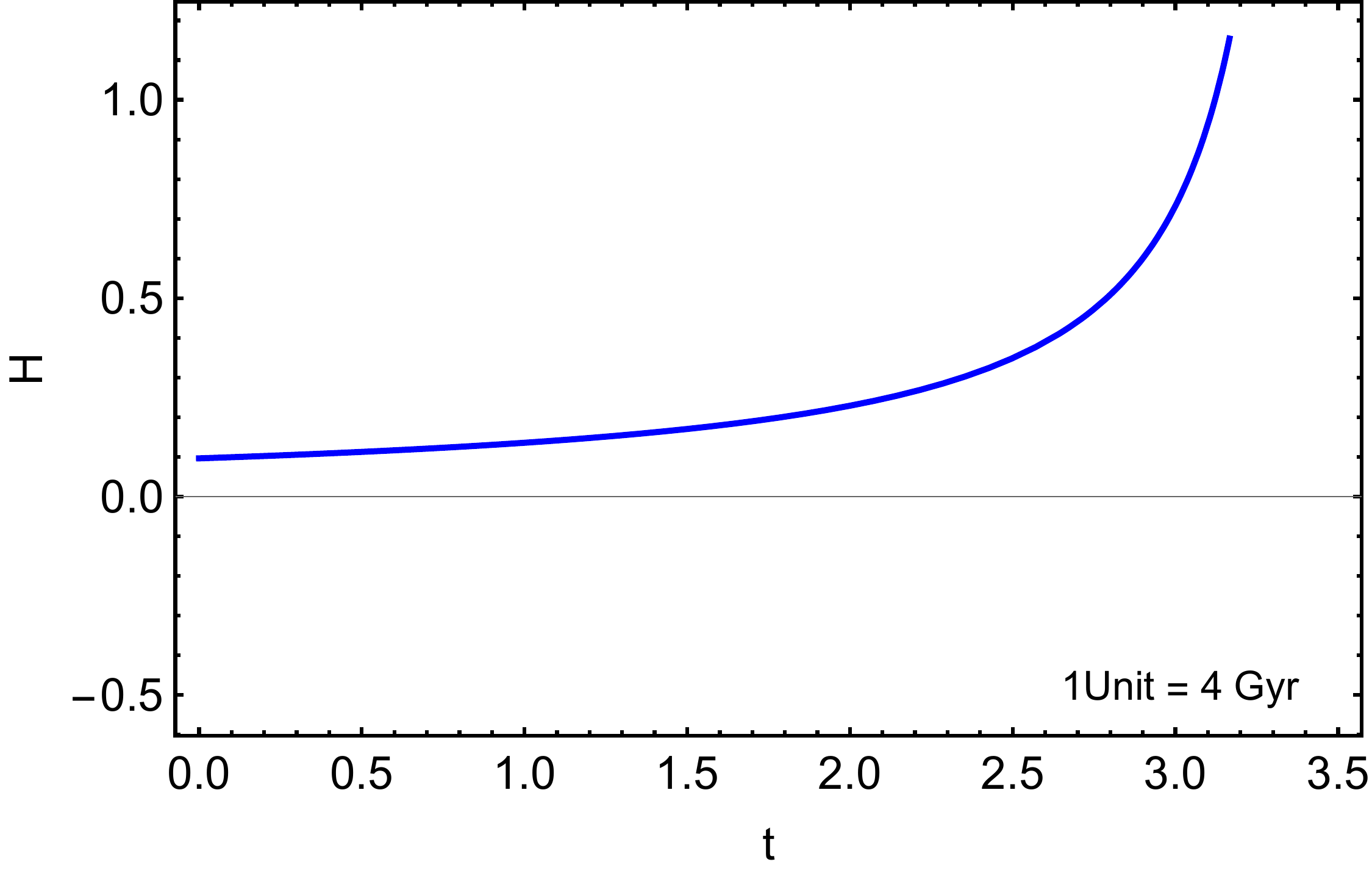,width=2.5in}}
\vspace*{8pt}
\caption{Variation of Hubble's parameter $H$ with time \textit{t} when $ k=3.45497$.\label{fig1}}
\end{figure}

\begin{figure}[H]
\centerline{\psfig{file=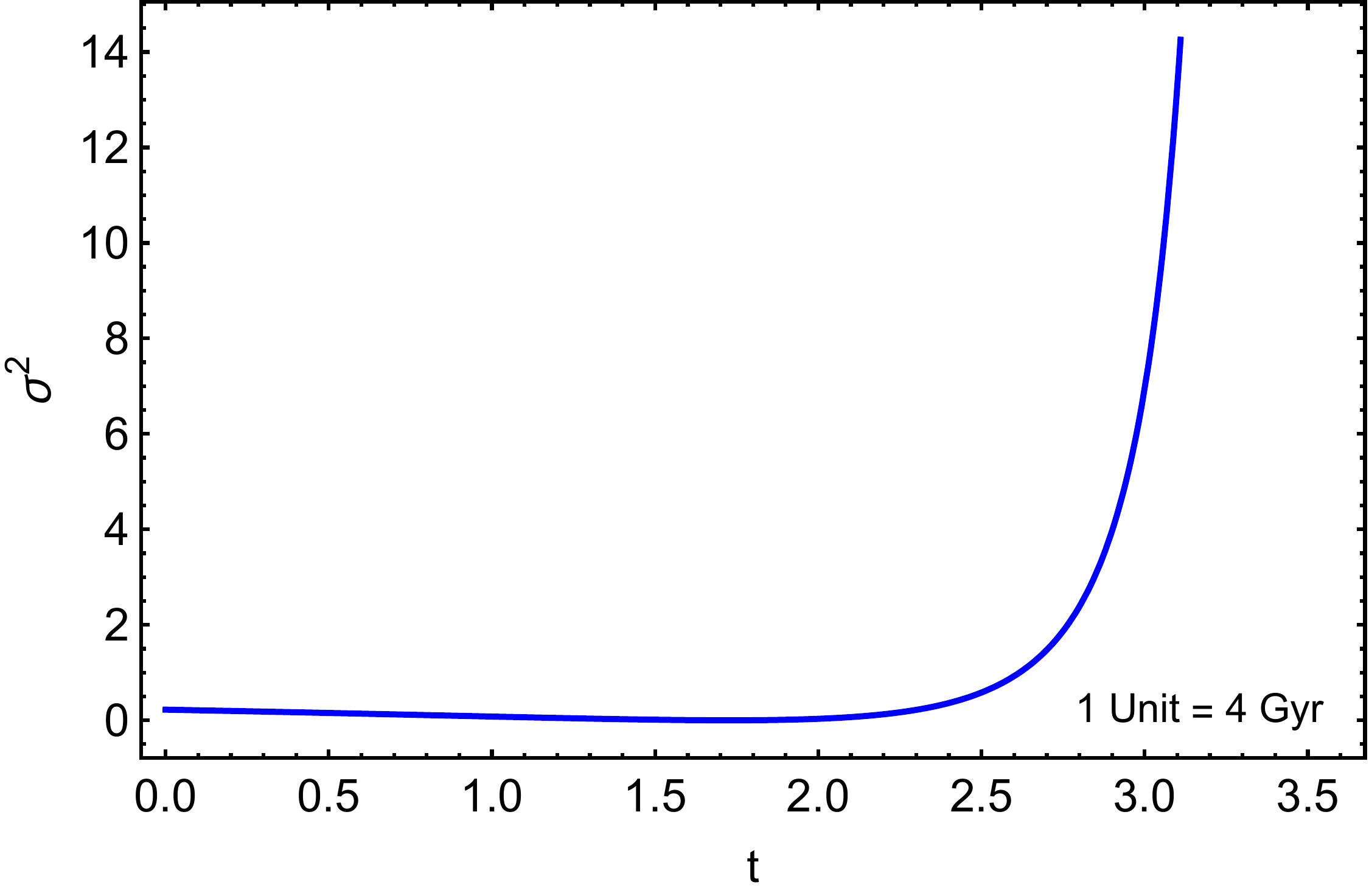,width=2.5in}}
\vspace*{8pt}
\caption{Variation of shear scalar $\sigma^{2}$ with time \textit{t} when $l=m=1, k=3.45497$.\label{fig1}}
\end{figure}

Accelerated expansion can be attained when $-1<q<0$ whereas $q<-1$ causes super-exponential expansion \cite{96}.  Figure 6 shows that the deceleration parameter $q$ is a negative constant -4 all through indicating that the model universe undergoes super-exponential expansion in the entire course of evolution. It may be noted that in higher dimensional theory with cosmological constant, super-exponential inflation (expansion) can be attained if $H$ increases with $t$ \cite{97,98,99}. In our case, $H$ is incresaing as shown in Fig. 7. Shear scalar $\sigma^{2}$ provides us the rate of deformation of the matter flow within the massive cosmos \cite{100}. From Fig. 8, we can see that $\sigma^{2}$ evolves almost constantly, then diverges after some finite time. From Eq. (19), the anisotropic parameter $A_h=0$. From these, we can sum up that initially, the isotropic universe expands with a slow and uniform change of shape, but after some finite time, the change becomes faster. 

Lastly, with a scale of 1 Unit = 4 Gyr, the point $t=3.45$ corresponds to 13.8 Gyr which align with $13.825\pm0.037$ Gyr, the present age of the universe estimated by the latest Planck 2018 result \cite{36}. At the point  $t=3.45$ and assuming $k=3.45497$, from Eq. (17), the numeric value of the Hubble's parameter is measured to be $H=67.0691$ which is very close to $H_0=67.36\pm0.54kms^{-1}Mpc^{-1}$, the value estimated by the latest Planck 2018 result \cite{36}.  

\section{Conclusion}

With due consideration of reasonable cosmological assumptions within the limit of the present cosmological scenario, we have analysed a spherically symmetric metric in 5D setting within the framework of LM. The model universe is predicted to be a DE model, dominated by vacuum energy. The displacement vector also acts as the time dependent DE. The model  represents an oscillating model, each cycle evolving with a big bang and ending at a big crunch, undergoing a series of bounces. Our universe undergoes super-exponential expansion in the entire course of evolution. Initially, the isotropic universe expands with a slow and uniform change of shape, but after some finite time, the change becomes faster. Then, the change slows down and tends to become uniform after expanding without any deformation of the matter flow for a finite time period. Lastly, the Hubble's parameter is measured to be $H=67.0691$ which is very close to $H_0=67.36\pm0.54kms^{-1}Mpc^{-1}$, the value estimated by the latest Planck 2018 result \cite{36}. We have constructed a model in LM appearing as a DE model; nonetheless, the work we have put forward is just a toy model. The model needs further deep study considering all the observational findings, which will be our upcoming work.

\end{document}